\documentclass[%
 aip,
 amsmath,amssymb,
 reprint,%
]{revtex4-1}

\usepackage{graphicx}
\usepackage{dcolumn}
\usepackage{bm}

\usepackage[utf8]{inputenc}
\usepackage[T1]{fontenc}
\usepackage{mathptmx}
\usepackage{etoolbox}
\usepackage{hyperref}
\usepackage{diagbox}
\usepackage{makecell}
\makeatletter
\def\@email#1#2{%
 \endgroup
 \patchcmd{\titleblock@produce}
  {\frontmatter@RRAPformat}
  {\frontmatter@RRAPformat{\produce@RRAP{*#1\href{mailto:#2}{#2}}}\frontmatter@RRAPformat}
  {}{}
}%
\makeatother
\begin{document}

\preprint{AIP/123-QED}

\title[]{Ultrafast manipulation of magnetic skyrmions by microwave fields}

\author{Xingdi Wang}
\affiliation{School of Materials and Physics, China University of Mining and Technology, Xuzhou 221116, P. R. China}

\author{Haiming Dong$^*$}
\email{hmdong@cumt.edu.cn}
\affiliation{School of Materials and Physics, China University of Mining and Technology, Xuzhou 221116, P. R. China}

\author{Kai Chang$^*$}
\email{kchang@zju.edu.cn}
\affiliation{Center for Quantum Matter, School of Physics, Zhejiang University, Hangzhou 310027, P. R. China}

\date{\today}

\begin{abstract}
We theoretically investigate the inertial dynamics of magnetic skyrmions driven by circularly polarized microwave-induced inverse Faraday effect (MIFE). By incorporating an inertial mass term into the Thiele equation and analytically deriving the microwave-induced magnetic fields and forces, we demonstrate fundamentally distinct dynamical regimes under continuous-wave (CW) versus pulsed excitation. Skyrmion inertia qualitatively transforms trajectories from smooth spirals to polygonal orbits under continuous driving, while enabling sustained post-pulse gyration that reveals the system's intrinsic relaxation dynamics. The handedness of the trajectory is determined by the topological charge and circularly polarized microwave (CPM) helicity: a left-circularly polarized (LCP) CPM attracts skyrmions toward the beam center, while a right-circularly polarized (RCP) CPM repels them. Systematic parameter analysis reveals how Gilbert damping, the intensity and frequency of CPM, and skyrmion mass control the transition between oscillatory and overdamped dynamical phases. Our work identifies inertia, topological charge, and CPM helicity as essential factors in ultrafast skyrmion manipulation and proposes a novel method for designing topological spin textures.
\end{abstract}

\maketitle

\section{Introduction}
Magnetic skyrmions are topologically protected spin textures characterized by nanoscale size, high stability, and distinctive dynamical responses to external perturbations. These properties make them promising candidates for next-generation spintronic devices \cite{nagaosa2013topological}. The dynamics of skyrmions under external stimuli have been extensively studied, including the effects of current-induced spin torques \cite{Iwasaki2013}, electric fields \cite{Ba2021}, spin waves \cite{Rozsa_2020}, thermal gradients \cite{Wangth2020}, and more. However, most studies of skyrmion dynamics do not consider inertial effects; they typically assume that skyrmions respond instantaneously to applied forces. For example, Zang et al. treated skyrmions as massless particles in studies of collective behavior \cite{zang2011dynamics}. However, Moutafis et al. observed that skyrmion trajectories in a parabolic potential are pentagonal instead of circular \cite{moutafis2009dynamics}. Makhfudz et al. attributed this discrepancy to skyrmions possessing inertial mass, which enables energy storage as kinetic energy during acceleration \cite{makhfudz2012inertia}. Subsequent studies confirmed the role of this inertial effect. Moon et al. realized controllable polygonal skyrmion trajectories via inertial effects in external magnetic fields \cite{moon2014control}, and Martinez et al. further showed that inertial mass is tunable by electric currents \cite{martinez2017mass}. Notably, Büttner et al. experimentally confirmed the existence of skyrmion inertial mass by analyzing gigahertz (GHz) cyclotron dynamics, identifying the contributions of energy storage during motion and dipolar interactions \cite{buttner2015dynamics}. These findings indicate that inertial effects become significant when the dynamical timescales approach or exceed the system's intrinsic frequencies. Therefore, understanding inertial dynamics is essential for accurately describing skyrmion motion beyond the quasi-static regime.

Ultrafast control of magnetic textures by microwaves and optics has attracted significant attention as a promising approach to achieve rapid operation, minimal energy consumption, and contact-free manipulation \cite{kirilyuk2010ultrafast, Driving020403}. Various physical mechanisms are being explored for the GHz microwave control of magnetic textures, including the generation, annihilation, and manipulation of skyrmions through nonequilibrium thermal effects induced by GHz microwave pulses or vortex beams \cite{fujita2017ultrafast,gerlinger2021application,kovacs2025all}. Meanwhile, researchers have explored ultrafast dynamical control of skyrmions using GHz microwave vortex beams carrying orbital angular momentum (OAM) \cite{guan2023optically,yang2018photonic}, and investigated the detection of skyrmions in two-dimensional van der Waals magnetic materials through the magneto-optic Kerr effect (MOKE) \cite{cai2024topological}. 

One notable nonlinear optical-magnetic effect is the inverse Faraday effect (IFE), in which circularly polarized light induces an effective magnetization in both nonmagnetic and magnetic materials \cite{kirilyuk2010ultrafast}. This effect opens a novel pathway for CPM-controlled magnetic structures \cite{pershan1966theoretical,van1965optically}. Recently, the microwave-induced inverse Faraday effect (MIFE) has been proposed and has demonstrated significant research value \cite{maje2021}, noting that the intrinsic frequency of skyrmions generally lies in the GHz band \cite{makhfudz2012inertia}. This effect provides a non-thermal pathway for ultrafast manipulation of magnetic structures \cite{kimel2005}. Recent studies have shown that the magnetic field generated by the IFE can induce a skyrmionic topological structure in plasmonic nanostructures \cite{yang2025dark}, and the all-optical control of skyrmion breathing is experimentally realized \cite{titze2024}. Furthermore, the topological charge of magnetic skyrmions in frustrated magnets can be controlled using CPM, a phenomenon known as the topological IFE \cite{miyata2022topological}.

To date, the MIFE of skyrmions has not been studied, particularly skyrmions that incorporate inertial effects. This gap is critical, as a continuous drive sustains forced oscillations, whereas a pulsed drive probes intrinsic relaxation, leading to fundamentally distinct dynamical phases. Existing models, which often neglect inertia or assume quasi-static fields, are inadequate to capture this dichotomy \cite{Milroad2022}. Here, we develop a unified inertial theoretical model incorporating the space-time-dependent IFE force to bridge this gap. We demonstrate that the temporal GHz microwave drive profile and skyrmion inertia jointly determine the emergent dynamical regimes, ranging from forced polygonal orbits under continuous illumination to persistent post-pulse gyration, revealing fundamental aspects of GHz topological spintronics.

\section{Theoretical Model}

\begin{figure}[!h]
\centering
\includegraphics[width=0.9\linewidth]{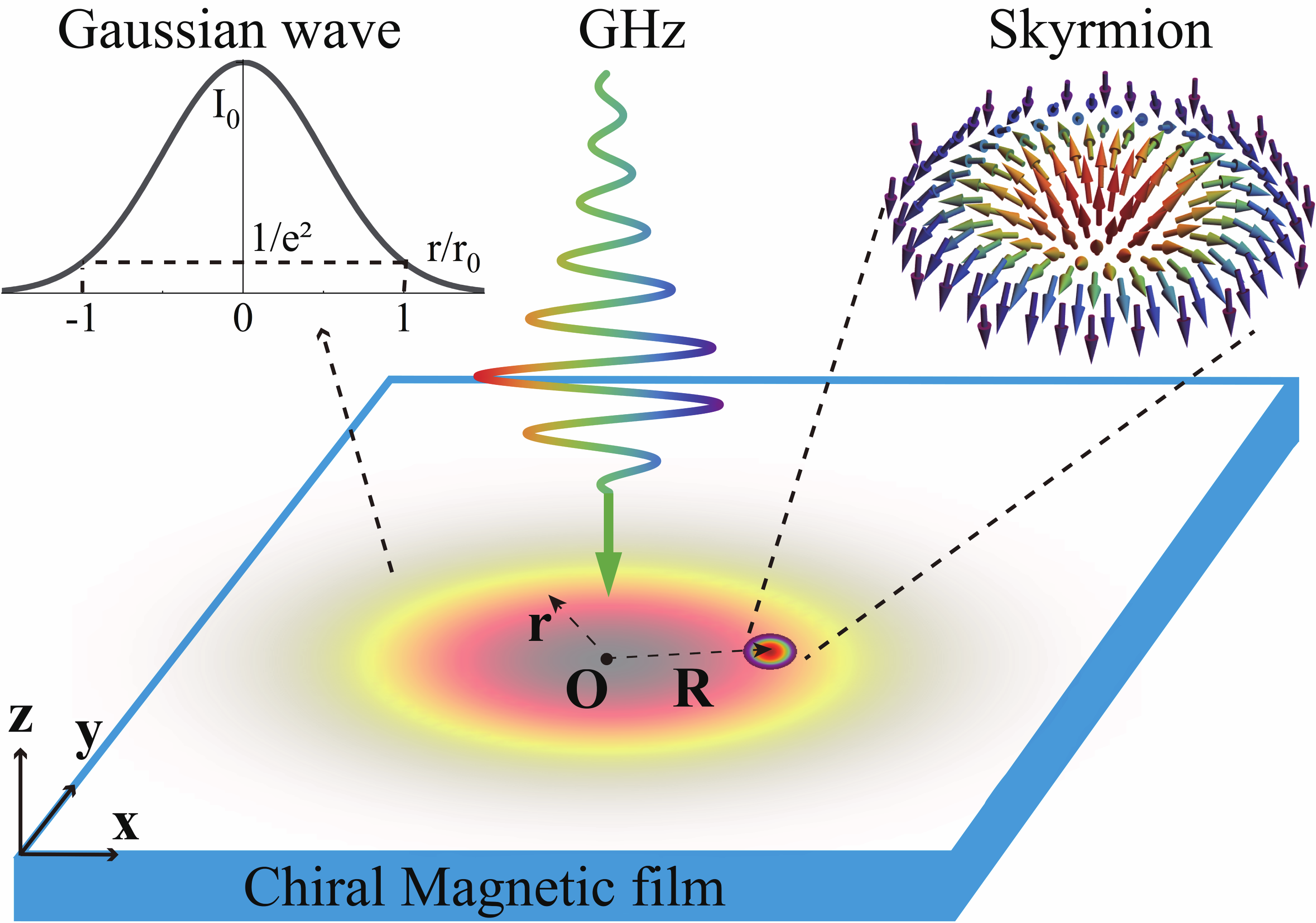}
\caption{Schematic diagram of Gaussian CPM driving Néel-type skyrmion motion in chiral magnetic film through MIFE. The origin of the Cartesian coordinate system ${\bf O}=(0, 0)$ is located at the center of the beam with $\boldsymbol{r} = (x, y)$. $\boldsymbol{R} = (X, Y)$ denotes the collective coordinate of a skyrmion. The inserts are the Gaussian beam profile (left) and a skyrmion configuration (right).}
\label{model}
\end{figure}

We consider a chiral magnetic ultrathin film, as shown in Fig. \ref{model}. The CPM interacts with magnetic structures in the film via IFE, such as skyrmions, to drive their dynamical evolution. For a CPM GHz wave propagating along the $+z$-axis as shown in Fig. \ref{model}, the electric field $\boldsymbol{E}(r,t)$ is given by
\begin{equation}
\boldsymbol{E}(r,t)=E_0(t) \exp \left( -\frac{r^2}{r_0^2} \right)(\boldsymbol{\hat{x}}\pm i\boldsymbol{\hat{y}}) \exp(-i \omega t),\label{E}
\end{equation}
where $E_0(t)$ denotes the peak electric field amplitude and $\omega$ is the GHz microwave angular frequency of a left-circularly polarized (LCP) and right-circularly polarized (RCP) Gaussian microwave fields, respectively. The GHz microwave intensity is related to the electric field by $I(r,t)=\frac{1}{2}c \epsilon_0 n |E(r,t)|^2$ and $I_0(t)=\frac{1}{2}c \epsilon_0 n |E_0(t)|^2$, with the speed $c$ of CPM in a vacuum, the vacuum permittivity $\epsilon_0$, and the refractive index $n$. We can obtain the magnetization $\boldsymbol{M}_{\text{\tiny{IFE}}}^{\text{\tiny{CW}}}$ via IFE under the CW excitation (see Appendix \ref{app:A} in detail), which reads
\begin{equation}     
\boldsymbol{M}_{\text{\tiny{IFE}}}^{\text{\tiny{CW}}}=\pm\frac{Ne^3}{2m_e^2}\frac{I(r)}{c \epsilon_0 n} \frac{\omega}{(\omega_0^2 - \omega^2)^2+(\beta   \omega)^2}\hat{\boldsymbol{z}},
\label{Mcw}
\end{equation}
for the RCP ($+$) and LCP ($-$)CPM fields, respectively. $\hat{\boldsymbol{z}}$ denotes the unit vector in the $z$-axis direction. $I(r,t)=I_0(t) \exp \left( -2r^2/r_0^2 \right)$. The IFE induces magnetization antiparallel to CPM propagation for LCP CPM and parallel for RCP CPM.

When the GHz microwave intensity is time-dependent, such as a Gaussian wave, we have $I_0(t)=I_0\exp \left[ -(t-t_0)^2/2\tau^2 \right]$ and $E_0(t)=E_0\exp \left[ -(t-t_0)^2/4\tau^2 \right]$. $\tau$ denotes the characteristic pulse width, and $t_0$ refers to the center time (peak time) of the pulse. The magnetization induced by RCP ($+$) and LCP ($-$) is, respectively, given by (see Appendix \ref{app:A})
\begin{equation}
\boldsymbol{M}_{\text{\tiny{IFE}}}^{\text{\tiny{Pulse}}}=\pm \frac{Ne^3}{2m_e^2} \frac{I(r,t)}{c \epsilon_0 n} \frac{\omega}{(\omega_0^2 - \omega^2)^2+(\beta   \omega)^2} \hat{\boldsymbol{z}},
\label{Mpul}
\end{equation}
accounting for the electron velocity-space distribution \cite{zhang2009simple}. The magnetization induced by the IFE has the same functional form for both CW and Gaussian pulse waves. The main difference is that the GHz microwave intensity is time-independent in the CW case, whereas it is time-dependent in the Gaussian pulse case. The effective magnetic field is given by
\begin{equation} 
{H}_{\text{\tiny{IFE}}}(\boldsymbol{r}) = \pm \mathcal{K}(\omega) I(r,t) \hat{\boldsymbol{z}},
\label{Hife}
\end{equation}
for RCP ($+$) and LCP ($-$) CPM, respectively. $\mathcal{K}(\omega)=\frac{N e^3}{2c \epsilon_0 n m_e^2 }\frac{\omega}{(\omega_0^2-\omega^2)^2+(\beta \omega)^2}$ is the frequency-dependent IFE coefficient. This effective magnetic field ${H}_{\text{\tiny{IFE}}}(\boldsymbol{r})$ is non-uniform in the transverse plane ($xy$-plane), thereby inducing a lateral force on the skyrmion.

As shown in Fig. \ref{model}, the normalized magnetization in the chiral magnetic film is given by $\boldsymbol{m}(\boldsymbol{r})=\boldsymbol{m}_0(\boldsymbol{r}-\boldsymbol{R})$. Here $\boldsymbol{R}$ denotes the collective coordinate of skyrmions as topologically protected quasiparticles. $\boldsymbol{r}$ is the position vector from any point on the $xy$ plane to the center of the CPM spot ${\bf O}=(0, 0)$. $\boldsymbol{m}_0(\boldsymbol{r}-\boldsymbol{R})$ represents the equilibrium magnetization texture. The magnetic potential energy of the system caused by the CPM field is
\begin{equation}
    U(\boldsymbol{R})=-\mu_0 M_s \int_V \boldsymbol{m}_0(\boldsymbol{r}-\boldsymbol{R}) \cdot \boldsymbol{H}_{\text{\tiny{IFE}}}(\boldsymbol{r}) dV,
\end{equation} 
by the effective magnetic field $\boldsymbol{H}_{\text{\tiny{IFE}}}(\boldsymbol{r})$. $\mu_0$ is the permeability of vacuum, and $M_s$ is the saturation magnetization of magnetic materials. The corresponding equivalent force is calculated by
\begin{equation}
\begin{aligned}       
\boldsymbol{F}_{\text{\tiny{IFE}}}&=-\nabla_{\boldsymbol{R}} U(\boldsymbol{R})\\
        &=-\mu_0 M_s \int_V \left( \frac{\partial m_z}{\partial x}\hat{\boldsymbol{x}}+\frac{\partial m_z}{\partial y}\hat{\boldsymbol{y}} \right){H}_{\text{\tiny{IFE}}}(\boldsymbol{r})dV.
    \end{aligned}
    \label{1}
\end{equation}

Since the beam spot size is significantly larger than the skyrmion diameter, the IFE magnetic field ${H}_{\text{\tiny{IFE}}}(\boldsymbol{r})$ can be Taylor-expanded about the skyrmion collective coordinate $\boldsymbol{R}$. Retaining terms up to the first order, it yields
${H}_{\text{\tiny{IFE}}}(\boldsymbol{r}) \approx {H}_{\text{\tiny{IFE}}}(\boldsymbol{R})+[(\boldsymbol{r}-\boldsymbol{R})\cdot \nabla]{H}_{\text{\tiny{IFE}}}(\boldsymbol{R})$.
We consider an axisymmetric Néel-type skyrmion in cylindrical coordinates in Fig. \ref{model}. The zeroth-order term in the expansion yields no net contribution upon spatial integration, and only the first-order term is retained. This indicates that a uniform CPM does not generate IFE forces for ideal skyrmions. We then introduce a cylindrical coordinate system $(\rho,\varphi,z)$ centered at the skyrmion position $\boldsymbol{R}$, where $\rho=|\boldsymbol{r}-\boldsymbol{R}|$ denotes the radial distance from the skyrmion center and $\varphi$ is the azimuthal angle, and we can obtain (see Appendix \ref{taylor})
\begin{equation}
    \boldsymbol{F}_{\text{\tiny{IFE}}}=-\mu_0 M_s d_0 \pi \nabla_{\perp} {H}_{\text{\tiny{IFE}}}(\boldsymbol{R}) \int \frac{\partial m_z}{\partial \rho}\rho^2 d\rho,
\label{Fint}
\end{equation}
where $\nabla_{\perp}$ is the in-plane gradient and $d_0$ is the effective thickness of a magnetic film. Taking into account a typical Néel-type skyrmion, the spin structure can be represented with the polar angle $\theta$ and the azimuthal angle $\varphi$ with $\boldsymbol{m}=(\sin \theta \cos \varphi,\sin \theta \sin \varphi, \cos \theta)$, we have
\begin{equation}
\nabla_{\perp} {H}_{\text{\tiny{IFE}}}(\boldsymbol{R})=-4\mathcal{K}(\omega) I_0(t) \frac{R}{r_0^2}\exp \left( -\frac{2R^2}{r_0^2} \right) \hat{\boldsymbol{r}},
\end{equation}
with the unit vector $\hat{\boldsymbol{r}}$ in the radial direction within the $xy$-plane and the integral in Eq.  \eqref{Fint}
\begin{equation}
\mathcal{I}=\int \frac{\partial m_z}{\partial \rho}\rho^2 d\rho=-\int \sin \theta \frac{\partial \theta}{\partial \rho}\rho^2 d\rho. \label{int}
\end{equation}
The spin texture of a skyrmion can be approximated by $\theta(\rho)=2 \arctan \left[ \sinh (R_d/r_d)/\sinh (\rho/r_d) \right]$, with $R_d=\pi D \sqrt{A/(16AK^2-\pi^2D^2K)}$ being the characteristic length scale associated with the skyrmion size and $r_d=\pi D/4K$ being the domain wall width \cite{braun1994fluctuations}. $A$ is the exchange constant, $D$ is the interfacial DMI constant, and $K = K_u - \mu_0 M_s^2 /2$ is the effective anisotropy constant, where $K_u$ is the perpendicular magneto-crystalline anisotropy in the magnetic film. By defining $R_d/r_d=u_0$ and $\rho/r_d=u$, we have (see Appendix \ref{app:eq10}) \cite{wang2018theory}
\begin{equation}
        \mathcal{I}=\int \frac{4 \sinh (u) \cosh (u) \sinh^2 (u_0)}{ [\sinh^2 (u_0) +\sinh^2 (u)]^2}r_d^2 u^2 du \approx 2R_d^2.
\label{integral}
\end{equation}

Finally, the complete expression for the IFE force is thus obtained as
\begin{equation}
\boldsymbol{F}_{\text{\tiny{IFE}}}=\pm 8\pi \mu_0 M_s d_0  \frac{R_d^2}{r_0^2} R\exp \left( -\frac{2R^2}{r_0^2} \right) \mathcal{K}(\omega) I_0(t) \hat{\boldsymbol {r}},
\label{IFE-F}
\end{equation}
for RCP ($+$) and LCP ($-$) microwaves, respectively. The expression $\boldsymbol{F}_{\text{\tiny{IFE}}}$ clearly indicates that, when exposed to non-uniform LCP, the IFE produces an effective attractive force that draws the skyrmion toward the beam center, while non-uniform RCP generates a repulsive force that pushes the skyrmion away from the beam center.

Skyrmions can be considered quasiparticles due to their topological protection properties and their center-of-mass dynamics. The dynamics of a magnetic skyrmion are described by the modified Thiele equation, which is \cite{buttner2015dynamics, makhfudz2012inertia, moon2014control}
\begin{equation}
M_{\text{sk}}\ddot{\boldsymbol R} + {\boldsymbol G} \times \dot{\boldsymbol R} + \alpha \mathcal{D} \dot{\boldsymbol R} = {\boldsymbol F}. \label{Thiele}
\end{equation}  
Here $M_{\text{sk}}$ denotes an effective inertial mass of a skyrmion, introduced phenomenologically to account for the delayed response of the internal magnetization texture during accelerated motion \cite{buttner2015dynamics}. $\dot{\boldsymbol R}=d{\boldsymbol R}/dt$ is the velocity of skyrmions. $\boldsymbol{G}=4\pi M_s d_0 Q \hat{\boldsymbol{z}}/\gamma$ is the gyrocoupling vector. The gyromagnetic ratio $\gamma=1.76 \times 10^{11}$A$\cdot$s$\cdot$kg$^{-1}$. $Q=\frac{1}{4\pi} \iint \boldsymbol{m} (\partial_x \boldsymbol{m} \times \partial_y \boldsymbol{m})dxdy$ is the topological charge. $\alpha$ is the Gilbert damping factor. $\mathcal{D}=4 \pi M_s d_0 \eta /\gamma$ is the dissipation tensor with $\eta$ being the shape factor of a skyrmion \cite{Liu2023Dynamics}. $\boldsymbol{F}$ is the force $\boldsymbol{F}_{\text{\tiny{IFE}}}$ induced by the IFE in this work.

The inertial term $M_{\text{sk}}\ddot{\boldsymbol{R}}$ in Eq.\eqref{Thiele} phenomenologically accounts for the delayed response of the spin texture during acceleration. Microscopically, this effective mass arises from the energy cost to deform the skyrmion's configuration and its intrinsic field structure \cite{buttner2015dynamics}. It can be estimated from field-theoretic considerations or micromagnetic simulations, typically yielding $M_{\text{sk}} \sim 10^{-23} - 10^{-22}$ kg for isolated skyrmions in thin films \cite{makhfudz2012inertia, paikaray2021skyrmion}. This mass introduces a characteristic timescale for velocity relaxation and enables second-order dynamics, fundamentally distinguishing it from the massless, first-order Thiele equation limit.

Additionally, without the skyrmion’s effective mass, the modified Thiele Eq. \eqref{Thiele} reduces to \cite{thiele1973steady}
\begin{equation}
    {\boldsymbol G} \times \dot{\boldsymbol R} + \alpha \mathcal{D} \dot{\boldsymbol R} = {\boldsymbol F}.
\end{equation}
We can obtain 
\begin{equation}
    \begin{aligned}
        \dot{R}_x&=\frac{\alpha D F_x+GF_y}{(\alpha D)^2+G^2},\\
        \dot{R}_y&=\frac{\alpha D F_y-GF_x}{(\alpha D)^2+G^2}.
    \end{aligned}
\end{equation}

We focus on the chiral magnetic films, i.e. Pt/Co/AlO$_x$ multilayer in this work to show our results, with the following parameters employed: the saturation magnetization $M_s=0.58$ MA$\cdot$m$^{-1}$, the exchange constant $A=15$ pJ$\cdot$m$^{-1}$, the interfacial DMI constant $D=3.5$ mJ$\cdot$m$^{-2}$, the perpendicular magneto-crystalline anisotropy $K_u=0.8$ MJ$\cdot$m$^{-3}$, the Gilbert damping factor $\alpha$ is approximately 0.01$-$0.1 and the effective thickness $d_0 = 1$ nm of magnetic Co film \cite{yang2018photonic,pizzini2014chirality}. Thus, the characteristic length scale $R_d=12.30$ nm and the domain wall width $r_d=4.66$ nm. $K=K_u - \mu_0 M_s^2/2 = 0.59$ MJ$\cdot$m$^{-3}$. Both the damping coefficient $\beta$ and the resonance frequency $\omega_0$ are set to zero in magnetic metals, which yields results consistent with quantum theory \cite{battiato2014quantum}. 

Moreover, the magnetization profile of an axisymmetric Néel type skyrmion can be further assumed to vary linearly with radius, the shape factor is given by $\eta=\frac{1}{4\pi} \iint (\partial \boldsymbol{m}/\partial x)\cdot (\partial \boldsymbol{m}/\partial x)dxdy=2\pi^2 R_d/8r_{dw} \approx 6.01$ with $r_{dw}=\sqrt{A/K} \approx 5.05$ nm \cite{jiang2017direct}. The Gaussian GHz microwave beam, with a waist radius of $r_0=100$ nm, has an intensity of approximately $10^6-10^7$ W/m$^2$ and operates at approximately 10 GHz. This frequency range is chosen to be comparable to the typical gyrotropic frequency of magnetic skyrmions on the order of GHz, placing the dynamics in a regime where inertial effects are pronounced and resonant responses can be explored. Practical excitation at such wavelengths can be achieved via plasmonic nanostructures that confine far-field radiation to sub-wavelength spots \cite{heeres2014subwavelength}. Plasmonic technologies allow for the beam waist to be smaller than the diffraction limit, achieving dimensions significantly smaller than the wavelength of the incident CPM \cite{heeres2014subwavelength, yin2018microwave, gorodetski2008observation, barnes2003surface}. The pulse duration is $\tau = 1$ ns, and the central time of the Gaussian pulse is $t_0 = 5$ ns. In the GHz frequency range, the complex refractive index of cobalt (Co) can be calculated as $\Tilde{n}= n+ \mathrm{i} \kappa = \sqrt{\Tilde{\epsilon_r}}$. $\kappa$ is the extinction coefficient, and $\Tilde{\epsilon_r}$ is the relative complex permittivity. The permittivity is often dominated by the Drude response $\Tilde{\epsilon_r} \approx \mathrm{i} \sigma_0/\epsilon_0 \omega$ \cite{basov2011electrodynamics}, with the conductivity $\sigma_0=1/\varrho \approx 1.60 \times 10^7$ S/m and $\varrho \approx 6.24 \times 10^{-8}$ $\Omega \cdot$m being the resistivity of Co \cite{matula1979electrical}. Accounting for finite-thickness and interfacial scattering effects in the Co thin film, the actual refractive index is lower than that of the bulk material \cite{fuchs1938conductivity, sondheimer2001mean}. 

The magnetic atomic number density of Co is $N_{\text{Co}}=\rho_{_{\text{Co}}} N_{A}/M_{\text{Co}} \approx 9.10 \times 10^{28}$ m$^{-3}$ with $\rho_{_{\text{\text{Co}}}} \approx 8.90$ g$\cdot$cm$^{-3}$ being the mass density of Co \cite{abdullaev2021density}, $M_{\text{Co}} \approx 58.93$ g$\cdot$mol$^{-1}$ being the molar mass of Co \cite{prohaska2022standard}, and $N_{A} \approx 6.02 \times 10^{23}$ mol$^{-1}$ being Avogadro's constant \cite{newell2019international}. Given that each Co atom contributes approximately 1-2 valence electrons, the free electron number density is estimated as $N_e \approx 0.91 - 1.82 \times 10^{29}$ m$^{-3}$ \cite{kittel2018introduction}. Accounting for finite-size effects and surface scattering in the thin film, the effective electron density is reduced relative to its bulk value. We adopt $N_e = 8 \times 10^{28}$ m$^{-3}$ as a representative value, which does not affect the qualitative nature of the underlying physical mechanisms. The effective mass $M_{\text{sk}}$ of skyrmions employed in this work is of the classical magnitude, approximately on the order of $10^{-23}$ kg \cite{makhfudz2012inertia,martinez2017mass,paikaray2021skyrmion}. 

\section{Results and Discussions}
\begin{figure}[htbp]
\centering
\includegraphics[width=0.9\linewidth]{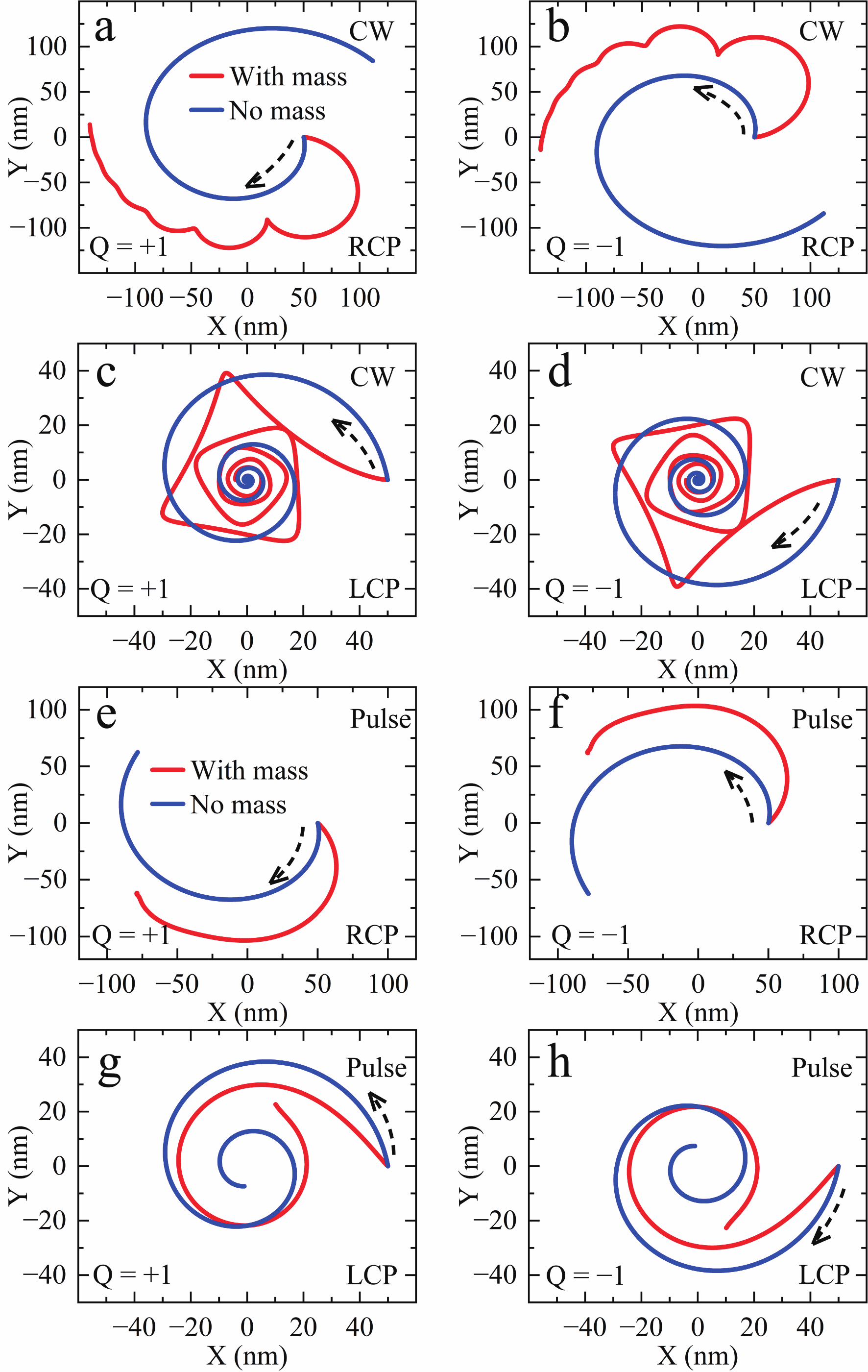}
\caption{The trajectories of skyrmions with $Q=\pm 1$ for the RCP/LCP CPM: (a)-(d) for CW CPM, and (e)-(h) for pulsed CPM, respectively. The skyrmions with mass (red) and without mass (blue) are included. The initial positions of skyrmions are all set to be 50 nm from the CPM beam center. $\alpha=0.03$, and $t=10$ ns for the motion time of skyrmions. The black arrows indicate the motion of the skyrmion.}
\label{trajectories}
\end{figure}

As illustrated in Fig. \ref{model}, a CPM beam with a Gaussian intensity profile is used to irradiate skyrmions in a chiral magnetic thin film. Due to the IFE, CPM fields induce an effective magnetic field $\boldsymbol{H}_{\text{\tiny{IFE}}}$ (see Eq.  \eqref{Hife}). The direction of this field depends on the CPM helicity, i.e., RCP or LCP. The spatial gradient of this non-uniform $\boldsymbol{H}_{\text{\tiny{IFE}}}$ creates an IFE force $\boldsymbol{F}_{\text{\tiny{IFE}}}$ on the magnetic skyrmion (as shown in Eq.  \eqref{IFE-F}), thereby propelling its motion in a specific direction. Skyrmions can be attracted to or repelled from the beam center, depending on the CPM helicity, as shown in Fig. \ref{trajectories}.

\begin{figure}[htbp]
\centering
\includegraphics[width=0.9\linewidth]{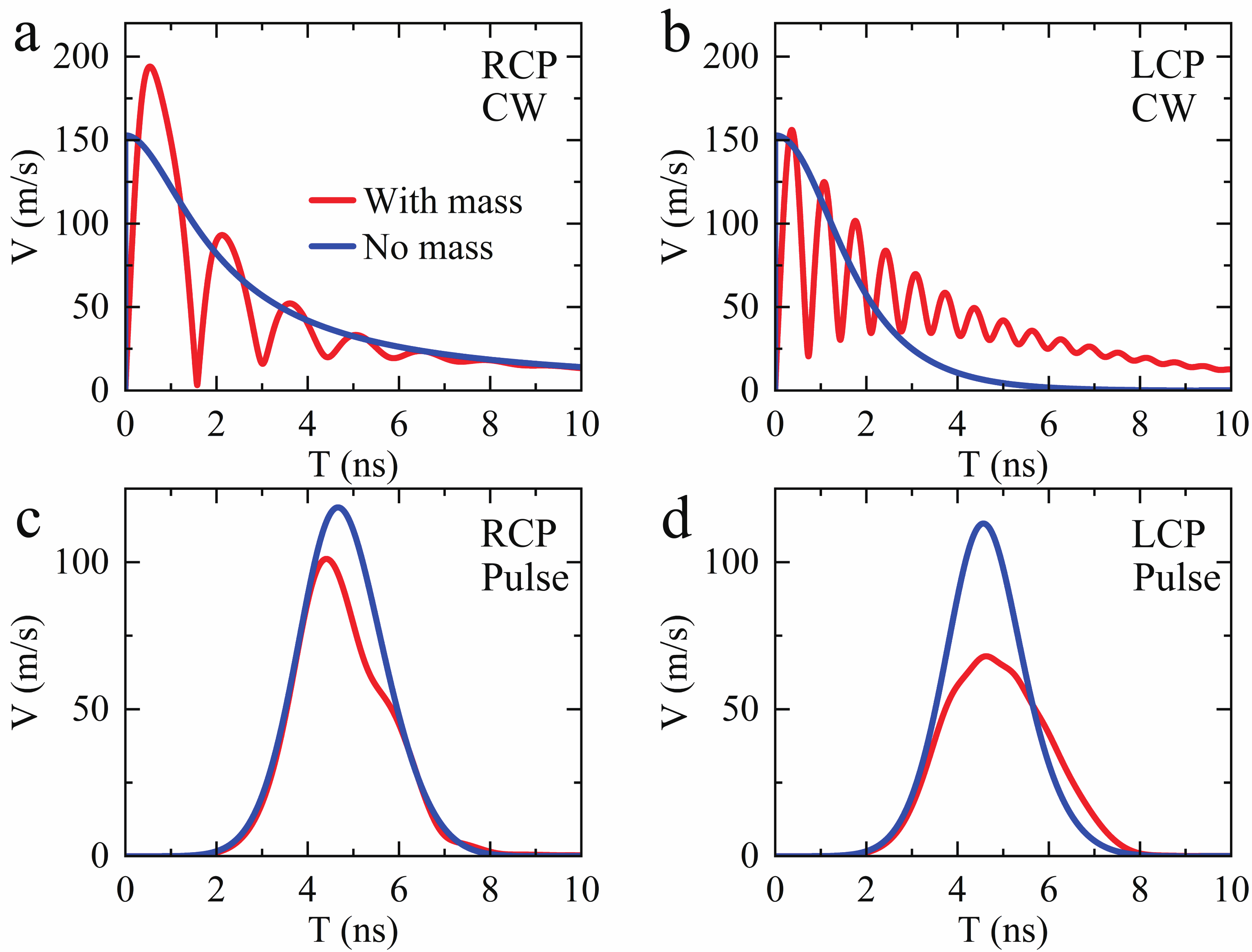}
\caption{Velocity-time dynamics of the magnetic skyrmion under RCP/LCP CPM. (a) and (b) for CW CPM, and (c) and (d) for pulsed CPM, respectively.}
\label{Velocity-time}
\end{figure}

Using the modified Thiele equation, we present numerical results on the dynamics of CPM-driven skyrmions. In this work, the skyrmion is treated as a massive quasiparticle influenced by gyrotropic, dissipative, inertial, and MIFE forces. The inclusion of the inertial term significantly changes the temporal response of the skyrmion's velocity and, as a result, its trajectory in real space. Fig. \ref{trajectories} illustrates the real-space trajectories of the skyrmion under CW and pulsed GHz microwave excitation, respectively. Moreover, the corresponding time evolutions of the skyrmion's velocity are shown in Fig. \ref{Velocity-time}. Together, these figures establish a direct connection between the temporal velocity response and the resulting spatial motion, providing a clear visualization of how the CPM field's helicity, topological properties, and inertial effects control skyrmion dynamics.

When a CW GHz microwave is applied in Fig. \ref{trajectories}(a-d), the trajectory of a skyrmion, without considering mass, follows a smooth spiral path. Its velocity gradually decreases from an initial maximum to a steady-state value, following the conventional Thiele equation, as shown in Fig. \ref{Velocity-time}(a) and (b). This equation indicates that the skyrmion's velocity is determined by the instantaneous balance between the GHz microwave driving force, the gyrotropic force, and dissipative forces. In this case, the system behaves as a first-order dynamical system, characterized by non-oscillatory trajectories (see blue lines in Fig. \ref{trajectories}). However, when inertial effects are taken into account, the trajectory changes significantly. As illustrated by the red curves, the skyrmion motion exhibits oscillatory characteristics, featuring petal-like and polygonal shapes that overlay the overall drift (see red lines in Fig. \ref{trajectories}). This behavior results from the inertial term in the modified Thiele equation, which shifts the system dynamics from first-order to second-order. As a consequence, the interplay between inertial and gyrotropic terms under a constant GHz microwave driving force induces cyclotron-like motion, preventing the system from smoothly following the external field. As a result, the velocity response shows damped oscillations before gradually settling into a steady state (see Fig. \ref{Velocity-time}).

Under pulsed CPM excitation in Fig. \ref{trajectories}(e-h), the skyrmion responds only during the GHz microwave pulse in the absence of inertia (mass), and its velocity waveform closely follows the temporal envelope of the CPM pulse. Once the GHz microwave driving force (the end of the pulse) ceases, the skyrmion’s motion stops immediately, resulting in short, non-oscillatory trajectories. However, when inertial effects are taken into account, the dynamics transform fundamentally. The peak velocity shifts significantly, exhibiting asymmetric rise and fall timescales (see Fig. \ref{Velocity-time}(c) and (d)). By analyzing the velocity-time responses alongside the real-space trajectories, we establish a clear dynamical connection between skyrmion inertia and GHz-driven motion. The mass of skyrmions introduces oscillatory relaxation and inertia effects in the dynamics of the velocity. This can be observed as trajectory modulation during continuous driving and persistent motion under pulsed excitation. These characteristics enable GHz microwave methods for identifying and characterizing inertial effects in skyrmion systems.

The direction in which skyrmion trajectories rotate, either clockwise or counterclockwise, is changed by both the topological charge $Q$ and the helicity of CPM, as illustrated in Fig. \ref{trajectories}. When the topological charge is reversed, the trajectory's chirality reverses, but its overall shape remains unchanged. This phenomenon is related to the reversal of the gyrotropic coupling vector $\boldsymbol{G}$, which is directly proportional to the skyrmion's topological charge $Q$. On the other hand, switching the circular polarization changes the trajectory's handedness and alters its geometric shape. LCP CPM drives skyrmions towards the center of the beam, while RCP CPM pushes them outward. This mechanism determines skyrmion trajectories: LCP CPM drives skyrmions toward the beam center, leading to localized accumulation, whereas RCP CPM drives them radially outward, resulting in spatial separation. Moreover, with a fixed CPM helicity, the initial sense of rotation of the trajectory indicates the sign ($\pm$) of $Q$. The dependence of the motion trajectory of skyrmions on CPM helicity and topological charge $Q$ is shown in Table \ref{tabI}. 

\begin{figure}[!h]
\centering
\includegraphics[width=0.9\linewidth]{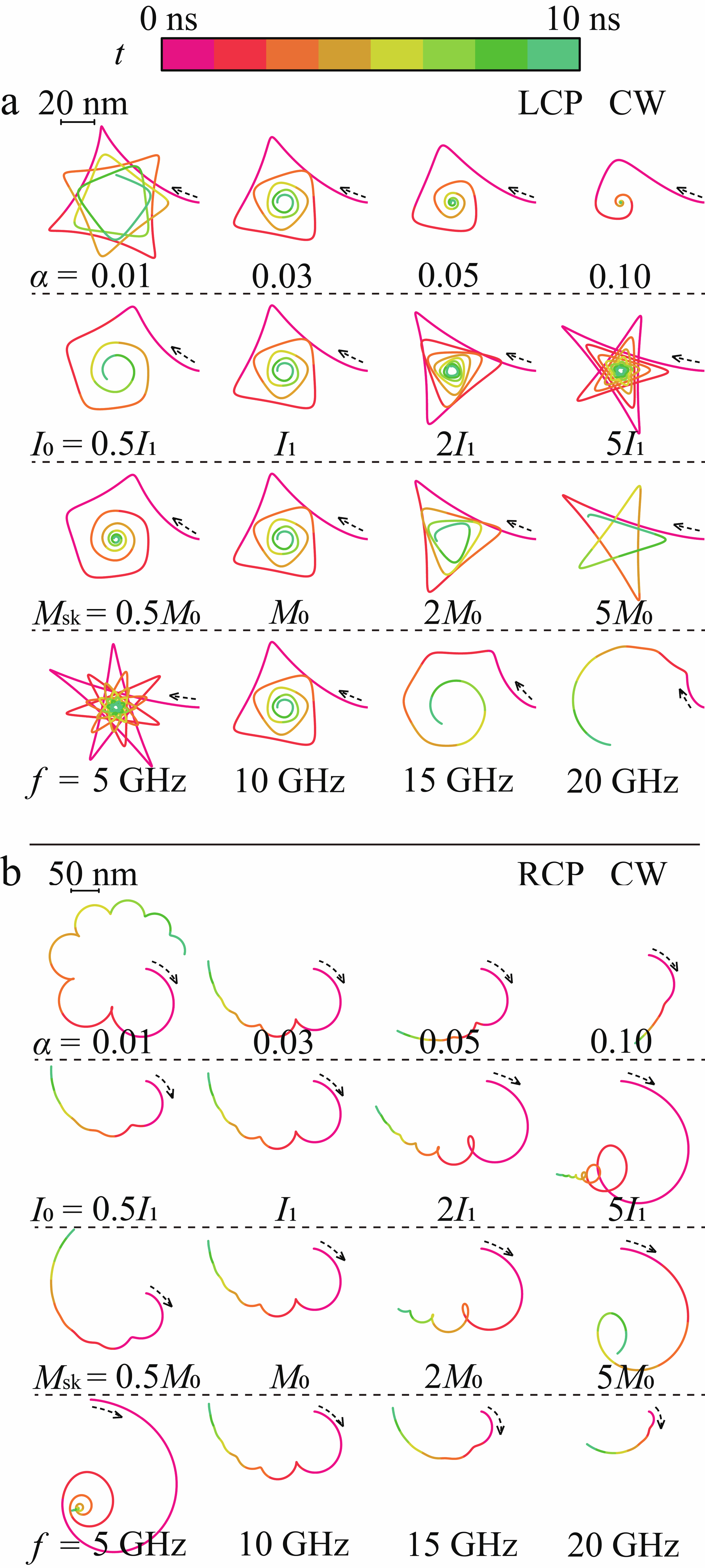}
\caption{Parameter-dependent trajectory morphology of inertial skyrmions under CW excitation with (a) LCP CPM and (b) RCP CPM, respectively. $\alpha=0.03$, $I_1=6.37 \times 10^6$ W/m$^2$, $M_{\text{sk}}=M_0=10^{-23}$ kg, $f=\omega/2\pi=10$ GHz. The temporal evolution of trajectories is represented by a color map (red to green: $0-10$ ns), while black arrows indicate the initial direction of motion. In all simulations, the skyrmion's initial position is fixed at 50 nm from the beam center, and its topological charge is $Q=+1$.}
\label{Parameter-trajectory-cw}
\end{figure}

\begin{table}[htb]
\centering
\caption{The dependence of the motion trajectory of skyrmions on CPM helicity and topological charge Q.}
\begin{tabular}{|c|c|c|c|}
\hline
\diagbox{Helicity }{Q} & $+$ & $-$  \\
\hline
RCP & \makecell[c]{Clockwise \\ Repulsion} & \makecell[c]{Counterclockwise\\Repulsion} \\
\hline
LCP & \makecell[c]{Counterclockwise \\ Attraction} & \makecell[c]{Clockwise\\ Attraction} \\   
\hline
\end{tabular}
\label{tabI}
\end{table}

\begin{figure}[!h]
\centering
\includegraphics[width=0.9\linewidth]{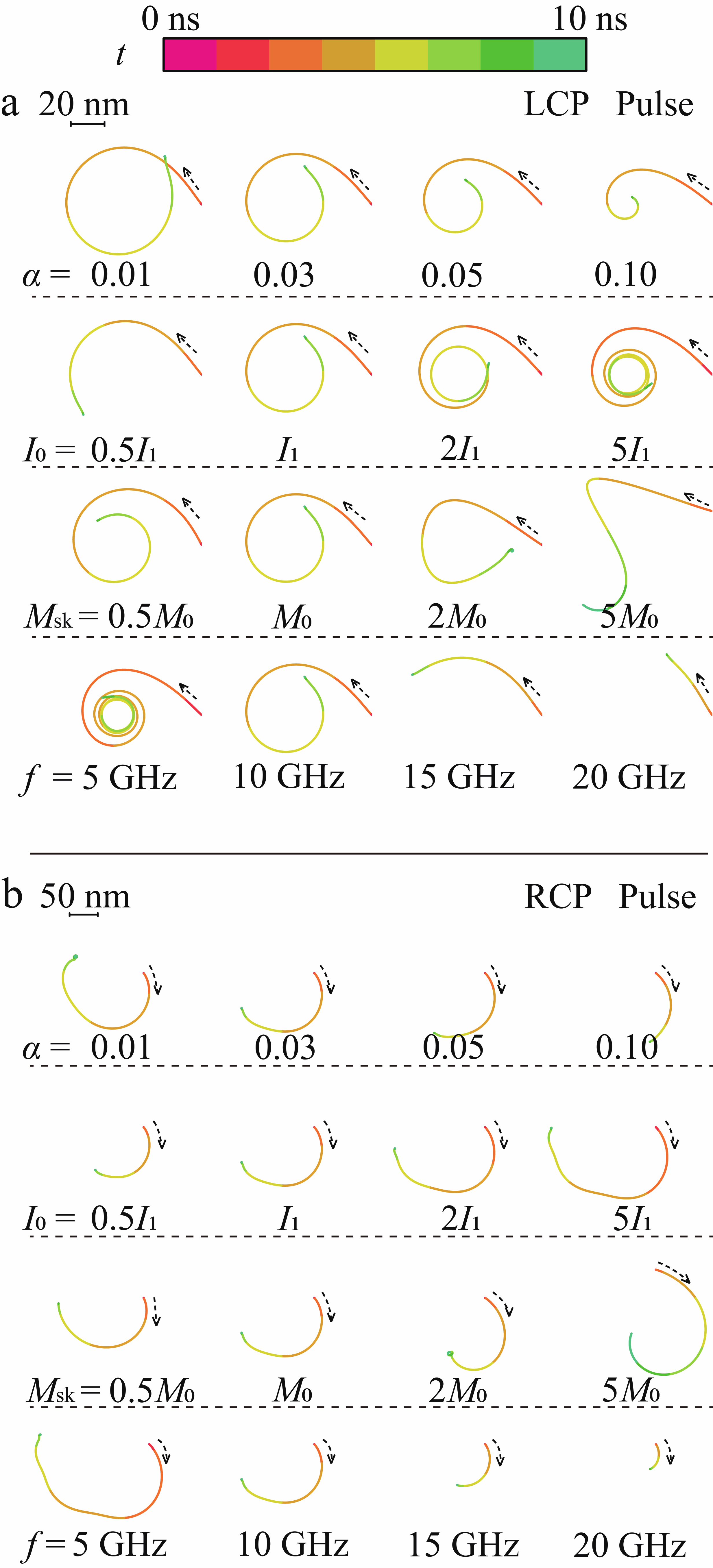}
\caption{Parameter-dependent trajectory morphology of inertial skyrmions under pulsed excitation for (a) LCP CPM and (b) RCP CPM, respectively, using consistent parameters and settings as shown in Fig. \ref{Parameter-trajectory-cw}.}
\label{Parameter-trajectory-pulsed}
\end{figure}

The analysis identifies inertia as a crucial factor in transforming skyrmion dynamics. It induces oscillatory relaxation and modulates trajectories, shifting the system from first- to second-order differential behavior. To clarify the underlying mechanism, we systematically investigate parameter dependencies. Figs. \ref{Parameter-trajectory-cw} and \ref{Parameter-trajectory-pulsed} illustrate the inertial skyrmion trajectories driven by CW and pulsed CPM, respectively. These trajectories are analyzed under systematic variations of the Gilbert damping factor $\alpha$, the GHz microwave intensity $I_0$, the effective mass $M_{\text{sk}}$, and the GHz microwave frequency $f=\omega/2\pi$. Despite the apparent diversity in trajectory shapes, all observed behaviors can be consistently explained by the modified Thiele equation that incorporates inertia.

As the damping coefficient increases, the trajectories of skyrmions under CW illumination change from highly oscillatory, star or petal-like paths to smoother, more compact spirals (see Fig. \ref{Parameter-trajectory-cw}). This change increases the dissipation of kinetic energy, thereby suppressing inertial oscillations. In the pulsed regime, higher damping quickly reduces the post-pulse motion (as indicated by the green section of the trajectory), thereby decreasing the spatial extent of the resultant trajectories (see Fig. \ref{Parameter-trajectory-pulsed}). An increase in Gilbert damping increases the dissipative term in the Thiele equation, leading to faster decay of inertial motion and suppression of extended polygonal trajectories.

Increasing the GHz microwave intensity $I_0$ increases the magnitude of the driving force $\boldsymbol{F}_{\text{\tiny{IFE}}}$ of Eq. \ref{IFE-F} in the modified Thiele Eq.  \eqref{Thiele}, leading to more pronounced trajectory deformation. Under CW GHz microwave excitation, a strong driving force amplifies inertial oscillations, leading to complex trajectories resembling petals or stars, as shown in Fig. \ref{Parameter-trajectory-cw}. For pulsed GHz microwave excitation, higher-intensity injection imparts greater momentum to the skyrmion, leading to a larger spiral trajectory, as illustrated in Fig. \ref{Parameter-trajectory-pulsed}. The inertial mass $M_{\text{sk}}$ primarily dictates the characteristic timescale of skyrmion dynamics. With a smaller mass, inertial oscillations are weak, and the trajectories resemble those of an overdamped system. As the mass increases, the coupling between the driving field and inertial motion becomes stronger. This leads to more pronounced oscillatory trajectories under continuous illumination and significantly enhanced residual motion following pulsed excitation.

The GHz microwave frequency $f$ is crucial in determining the dynamical regime of skyrmions. At low frequencies, the driving force is quasi-adiabatic, allowing the full development of inertial oscillations, resulting in complex trajectories that resemble star-like or multi-loop patterns. Conversely, at sufficiently high frequencies, the skyrmion cannot keep pace with the rapid field changes. This results in limited displacement and smoother trajectories. Both the GHz microwave frequency and CPM intensity play key roles in the driving force term of the modified Thiele equation, as illustrated in Figs. \ref{Parameter-trajectory-cw} and \ref{Parameter-trajectory-pulsed}.

Additionally, under pulsed GHz microwave excitation in Fig. \ref{Parameter-trajectory-pulsed}, skyrmion dynamics can be clearly divided into two stages: a momentum-injection stage followed by an inertial-relaxation stage. While increasing the GHz microwave intensity enhances the skyrmion's instantaneous motion during the pulse, it has minimal impact on the trajectory after the pulse ends. In contrast, a larger inertial mass significantly prolongs the motion after the pulse terminates, suggesting that post-pulse dynamics are primarily driven by inertial relaxation rather than by external driving forces. This behavior underscores the distinct roles of GHz microwave intensity in injecting momentum and inertial mass in retaining momentum.

A comparison between CW and pulsed GHz microwave excitation indicates the importance of temporal driving, as illustrated in Figs. \ref{Parameter-trajectory-cw} and \ref{Parameter-trajectory-pulsed}. CW CPM fields sustain forced inertial oscillations, yielding trajectory shapes highly sensitive to damping, mass, and frequency. In contrast, pulsed CPM excitation reveals the skyrmion's natural inertial response: the trajectories after a pulse directly reflect the stored kinetic energy and its dissipation, providing clear insight into inertial effects. Under CW conditions, the solution remains in a steady state, with the trajectory shape influenced by the drive frequency, damping, and natural frequency. In contrast, under pulsed conditions, it reduces to an initial-value problem, in which the trajectory illustrates the excitation and decay of the system's natural modes. These findings demonstrate that the dynamics of inertial skyrmions under CPM can be effectively tuned by adjusting damping, GHz microwave intensity, inertial mass, and frequency.

\begin{figure}[!t]
\centering
\includegraphics[width=0.9\linewidth]{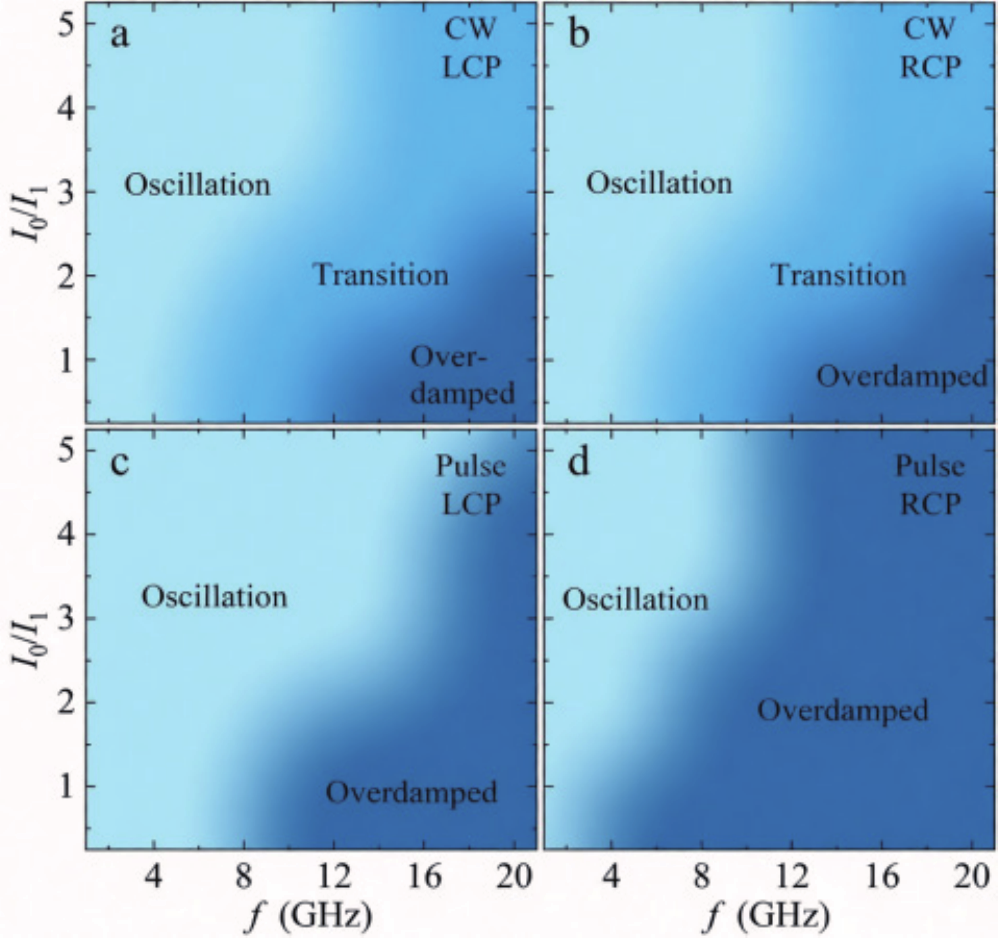}
\caption{Phase diagram of skyrmion dynamics by CPM with the intensity $I_0$ and the frequency $f=\omega/2\pi$ for (a) CW LCP, (b) CW RCP, (c) pulsed LCP, and (d) pulsed RCP CPM. $I_1=6.37 \times 10^6$ W/m$^2$.}
\label{phase}
\end{figure}

The phase diagrams presented in Fig. \ref{phase} serve as a systematic characterization of the inertial dynamics explored in this work. They map the emergent dynamical regimes, oscillatory, transition, and overdamped, across the two-dimensional parameter space of GHz microwave intensity $I_0$ and frequency $f$, for both CW and pulsed excitation with LCP and RCP CPM, respectively.
These regimes can be quantified by identifying two key parameters in the inertial Thiele equation \eqref{Thiele}. The first is a driving strength ratio $\beta \propto I_0 / f^3$, derived from the force scaling $\boldsymbol{F}_{\text{\tiny{IFE}}} \propto I_0/f^3$ (Eq. \ref{IFE-F}). The second is a relative frequency $\zeta = f / f_{\text{gy}}$, where $f_{\text{gy}} \propto |G|/M_{\text{sk}}$ is the characteristic gyration frequency of the inertial skyrmion. The oscillatory phase for high $\beta$ and $\zeta \sim 1$ corresponds to resonant or strongly driven conditions where inertia dominates the response. The overdamped phase emerges when $\beta$ is sufficiently small or when $\zeta \gg 1$ with moderate-to-low $\beta$. The GHz microwave driving is either too weak or too rapid for the skyrmion to follow, leaving dissipation as the governing factor. The transition phase is a hallmark of sustained CW driving, arising from the off-resonant interplay between the continuous external drive and the system’s natural oscillatory mode.

An analysis of this synthesis reveals the critical role of the temporal drive profile. Under CW illumination, the persistent force allows the system to settle into a steady state dictated by the balance of drive, inertia, and damping, making all three phases accessible. In stark contrast, pulsed illumination acts as a transient perturbation. The subsequent dynamics are primarily a free relaxation of the intrinsically excited inertial mode, which explains the absence of a well-defined transition phase and the stark bifurcation into either decaying gyration (oscillatory) or immediate quenching (overdamped). This difference demonstrates the conceptual distinction between forced and intrinsic inertial dynamics discussed throughout this work. Finally, the phase diagrams encode the effect of CPM helicity. The sign reversal in $\boldsymbol{F}_{\text{\tiny{IFE}}}$ of Eq. \ref{IFE-F} between LCP and RCP effectively inverts the local potential landscape near the beam center, creating a confining well for LCP and a repulsive barrier for RCP. This shifts the effective resonance condition and distorts the phase boundaries, revealing how CPM helicity jointly determines the inertial dynamics of GHz microwave-driven skyrmions.

\section{Conclusions}  

In summary, we have established a theoretical work that reveals inertia as an essential component linking the temporal structure of the CPM field and the dynamics of skyrmions. Our work identifies three principal findings: (i) It provides an analysis model based on the inertial Thiele equation with an IFE-derived force for ultrafast GHz manipulation; (ii) The trajectory, handedness, and spatial evolution are jointly controlled by the skyrmion's topological charge and the CPM's helicity, enabling selective attraction or repulsion relative to the beam center. It identifies the inertial effect, sustained post-pulse gyration, as a hallmark signature of skyrmion mass; and (iii) it demonstrates that Gilbert damping, GHz microwave intensity, frequency, and pulse duration form a parameter space for engineering skyrmion trajectories, from stabilized orbits to controlled relaxation. This clarifies the distinct physics of CW versus pulsed driving and provides a means to GHz-microwave manipulate skyrmions. We hope that this work can not only advance the theoretical understanding of inertial skyrmion dynamics but also provide a predictive foundation for the design of GHz spintronic devices based on topological magnetic textures.

\section*{ACKNOWLEDGMENTS}  This work is funded by the Science and Technology Program of Xuzhou (KC25001) and supported by the National Natural Science Foundation of China (Grant Nos. 12374079 and 11604380).

\section*{AUTHORDECLARATIONS}

Conflict of Interest

The authors have no conflicts to disclose.

\section*{Data Availability Statement}
The data that support the findings of this study are available from the corresponding author upon reasonable request.

\appendix
\renewcommand\thefigure{A1}

\section{Inverse Faraday Effect}\label{app:A}
The MIFE denotes the induction of net magnetization in materials by CPM. The magnetic moment generated via IFE is equivalent to an effective magnetic field that can interact with magnetic structures, such as skyrmions, to drive their dynamical evolution. With the wave spot center of CPM as shown in Fig. \ref{model}, the intensity profile of a Gaussian beam at the beam waist plane ($z$=0) is given by
\begin{equation}
    I(r,t)=I_0(t) \exp \left( -\frac{2r^2}{r_0^2} \right),\label{I}
\end{equation}
where $I_0(t)$ denotes the peak intensity at the beam center, depending on the time $t$. $r_0$ is the beam waist radius. $\boldsymbol{r}=(x, y)$ represents the radial coordinate. We find that a spatially non-uniform GHz microwave field can induce an effective magnetic field through the IFE, and this field gradient exerts an equivalent force on the skyrmion in chiral magnetic films (see Fig. \ref{model}).

The GHz microwave intensity is related to the electric field by $I(r,t)=\frac{1}{2}c \epsilon_0 n |E(r,t)|^2$ and $I_0(t)=\frac{1}{2}c \epsilon_0 n |E_0(t)|^2$. The electric force acting on electrons in the material is $\boldsymbol{F}_e=-e\boldsymbol{E}(r,t)$ with the elementary charge $e$. The equation of motion for electrons is described by the Drude-Lorentz model \cite{battiato2014quantum, sehmi2017optimizing}, 
\begin{equation}
\ddot{\boldsymbol{r}}_e+\beta \dot{\boldsymbol{r}}_e + \omega_0^2 \boldsymbol{r}_e=-\frac{e\boldsymbol{E}(r,t)}{m_e}, \label{DL}
\end{equation}
where $\boldsymbol{r}_e$ is the displacement of the electron from its equilibrium position. $\dot{\boldsymbol{r}}_e=d\boldsymbol{r}_e/dt$. $m_e$ is the effective mass of the electron, $\beta$ is the phenomenological damping coefficient, and $\omega_0$ is the natural resonance frequency of the bound electron system. The model is often used to describe electromagnetic properties of free and bound electrons \cite{app11219902}.

For the continuous-waves (CW), where the GHz microwave intensity is time-independent with $I_0(t)=I_0$ and $E_0(t)=E_0$, the solution of Eq.  \eqref{DL} is written as
\begin{equation}
    \boldsymbol{r}_e(r,t)= r_{e0}\exp \left(-\frac{r^2}{r_0^2}\right)(\hat{\boldsymbol{x}}\pm i\hat{\boldsymbol{y}})\exp(-i\omega t).
\end{equation}
Here $r_{e0} = -(eE_0/m_e)/[(\omega_0^2-\omega^2)-i\beta \omega]$ is the complex amplitude. The solution reveals that the electron undergoes helical motion driven by CPM, resulting in a looped current in the plane transverse to the propagation direction (the $xy$ plane). We consider the electron velocity-space distribution in magnetic metals with an effective electron density $N$, the resulting volume magnetization is given by $N\boldsymbol{\mu}_r$ with the magnetic moment $\boldsymbol{\mu}_r = -\frac{e}{2}(\boldsymbol{r}_e\times d\boldsymbol{r}_e/dt)$, and we obtain the magnetization $\boldsymbol{M}_{\text{\tiny{IFE}}}^{\text{\tiny{CW}}}$ via IFE under the CW excitation,
which reads
\begin{equation}     
\boldsymbol{M}_{\text{\tiny{IFE}}}^{\text{\tiny{CW}}}=\pm\frac{Ne^3}{2m_e^2}\frac{I(r)}{c \epsilon_0 n} \frac{\omega}{(\omega_0^2 - \omega^2)^2+(\beta \omega)^2}\hat{\boldsymbol{z}},
\label{Mcw}
\end{equation}
for the RCP and LCP  fields, respectively. 

When the GHz microwave intensity is time-dependent, such as a Gaussian pulse, we have $I_0(t)=I_0\exp \left[ -(t-t_0)^2/2\tau^2 \right]$ and $E_0(t)=E_0\exp \left[ -(t-t_0)^2/4\tau^2 \right]$. $\tau$ denotes the characteristic pulse duration (width), and $t_0$ is the center time (peak time) of the pulse. The assumed solution of Eq. (\ref{DL}) is
\begin{equation} \boldsymbol{r}_e(r,t)=r_{e0}(t)\exp \left(-\frac{r^2}{r_0^2}\right)(\hat{\boldsymbol{x}}\pm i\hat{\boldsymbol{y}})\exp(-i\omega t).
\end{equation}
Substituting the assumed solution back into the equation, we have
\begin{equation}
\begin{aligned}
    &\ddot{r}_{e0}(t)+(\beta-2i\omega)\dot{r}_{e0}(t)+(\omega_0^2-\omega^2-i\beta \omega)r_{e0}(t)\\&=-\frac{eE_0}{m_e}\exp \left[ -\frac{(t-t_0)^2}{4\tau^2} \right].
\end{aligned}
\end{equation}
The first- and second-order time derivatives of the displacement can be neglected under the instantaneous-response approximation, in which electrons respond instantaneously to the GHz microwave field. We then obtain the following equation,
\begin{equation}
    (\omega_0^2-\omega^2-i\beta \omega)r_{e0}(t)=-\frac{eE_0}{m_e}\exp \left[ -\frac{(t-t_0)^2}{4\tau^2} \right].
\end{equation}
By analyzing, we have
\begin{equation}
    r_{e0}(t)=r_{e0} \exp \left[ -\frac{(t-t_0)^2}{4\tau^2} \right].
\end{equation}
The magnetization induced by an RCP and an LCP Gaussian pulse is, respectively, given by
\begin{equation}
\boldsymbol{M}_{\text{\tiny{IFE}}}^{\text{\tiny{Pulse}}}=\pm \frac{Ne^3}{2m_e^2} \frac{I(r,t)}{c \epsilon_0 n} \frac{\omega}{(\omega_0^2 - \omega^2)^2+(\beta   \omega)^2} \hat{\boldsymbol{z}},
\label{Mpul}
\end{equation}
accounting for the electron velocity-space distribution. Finally, Eqs. \ref{Mcw} and \ref{Mpul} present the magnetization via IFE generated by CW and pulsed CPM in a magnet, respectively.

\section{Analysis and derivation for Eq.  \ref{Fint}}\label{taylor}
Due to the magnetic potential $U(\boldsymbol{R})$ caused by the CPM field, the IFE force is 
\begin{equation}
\begin{aligned}       
\boldsymbol{F}_{\text{\tiny{IFE}}}&=-\nabla_{\boldsymbol{R}} U(\boldsymbol{R})\\
        &=\mu_0 M_s \int_V\nabla_{\boldsymbol{R}}[\boldsymbol{m}_0(\boldsymbol{r}-\boldsymbol{R})\cdot \boldsymbol{H}_{\text{\tiny{IFE}}}(\boldsymbol{r})]dV,
    \end{aligned}  \label{FUR}
\end{equation}
with
\begin{equation}
    \begin{aligned}
    &~~~\nabla_{\boldsymbol{R}}[\boldsymbol{m}_0(\boldsymbol{r}-\boldsymbol{R})\cdot \boldsymbol{H}_{\text{\tiny{IFE}}}(\boldsymbol{r})]\\&=-\boldsymbol{H}_{\text{\tiny{IFE}}}(\boldsymbol{r}) \times [\nabla_{\boldsymbol{r}}\times \boldsymbol{m}_0(\boldsymbol{r}-\boldsymbol{R})]\\
        &-[\boldsymbol{H}_{\text{\tiny{IFE}}}(\boldsymbol{r})\cdot \nabla_{\boldsymbol{r}}]\boldsymbol{m}_0(\boldsymbol{r}-\boldsymbol{R}).
    \end{aligned}
\end{equation}
In this context, the operator $\nabla_{\boldsymbol{R}}$ acts only on $\boldsymbol{m}_0(\boldsymbol{r}-\boldsymbol{R})$, with $\nabla_{\boldsymbol{R}} = -\nabla_{\boldsymbol{r}}$. Additionally, the operation rule is given by $\nabla(\boldsymbol{A}\cdot \boldsymbol{B}) = \boldsymbol{A} \times (\nabla \times \boldsymbol{B}) + (\boldsymbol{A}\cdot \nabla)\boldsymbol{B} + \boldsymbol{B} \times (\nabla \times \boldsymbol{A}) + (\boldsymbol{B}\cdot \nabla)\boldsymbol{A}$ for any vectors $\boldsymbol{A}$ and $\boldsymbol{B}$.

Given that $\boldsymbol{H}_{\text{\tiny{IFE}}}(\boldsymbol{r})$ has only a component in the $z$ direction and that $\boldsymbol{m}_0(\boldsymbol{r}-\boldsymbol{R})$ is uniformly distributed along the $z$ axis, we have
\begin{equation}
    \begin{aligned}
        &~~~\boldsymbol{H}_{\text{\tiny{IFE}}}(\boldsymbol{r}) \times [\nabla_{\boldsymbol{r}}\times \boldsymbol{m}_0(\boldsymbol{r}-\boldsymbol{R})]\\&=\left( \frac{\partial m_z}{\partial x}\boldsymbol{\hat{x}}+\frac{\partial m_z}{\partial y}\boldsymbol{\hat{y}} \right)H_{\text{\tiny{IFE}}}(\boldsymbol{r}),
    \end{aligned}
\end{equation}
and
\begin{equation}
    [\boldsymbol{H}_{\text{\tiny{IFE}}}(\boldsymbol{r})\cdot \nabla_{\boldsymbol{r}}]\boldsymbol{m}_0(\boldsymbol{r}-\boldsymbol{R})=0.
\end{equation}
Thus, the Eq.  \ref{FUR} is written as
\begin{equation}
    \boldsymbol{F}_{\text{\tiny{IFE}}}=-\mu_0 M_s \int_V \left( \frac{\partial m_z}{\partial x}\boldsymbol{\hat{x}}+\frac{\partial m_z}{\partial y}\boldsymbol{\hat{y}} \right)H_{\text{\tiny{IFE}}}(\boldsymbol{r})dV.
\end{equation}

Since the beam spot size is significantly larger than the skyrmion diameter, the effective magnetic field due to the IFE ${H}_{\text{\tiny{IFE}}}(\boldsymbol{r})$ can be approximated using the Taylor expansion $\boldsymbol{R}$, retaining terms only up to the first order, which reads
\begin{equation}
    {H}_{\text{\tiny{IFE}}}(\boldsymbol{r}) \simeq {H}_{\text{\tiny{IFE}}}(\boldsymbol{R})+[(\boldsymbol{r}-\boldsymbol{R})\cdot \nabla]{H}_{\text{\tiny{IFE}}}(\boldsymbol{R}).
\end{equation}
We consider a Néel-type skyrmion. The zeroth-order term in the expansion contributes nothing upon spatial integration, while the first-order term is retained, which reads

\begin{equation}
\begin{aligned}
    &~~~[(\boldsymbol{r}-\boldsymbol{R})\cdot \nabla]{H}_{\text{\tiny{IFE}}}(\boldsymbol{R})\\&=\left[ (x-R_x)\frac{\partial}{\partial x}+(y-R_y)\frac{\partial}{\partial y} \right]{H}_{\text{\tiny{IFE}}}(\boldsymbol{R}).
\end{aligned}
\end{equation}
We introduce a cylindrical coordinate system $(\rho,\varphi,z)$ centered at the skyrmion position $\boldsymbol{R}$, where $\rho=|\boldsymbol{r}-\boldsymbol{R}|$ denotes the radial distance from the skyrmion center and $\varphi$ the azimuthal angle. Through the transformation relationship between cylindrical coordinates and rectangular coordinates, we obtain
\begin{equation}
    \boldsymbol{F}_{\text{\tiny{IFE}}}=-\mu_0 M_s t_0 \pi \nabla_{\perp} H_{\text{\tiny{IFE}}}(\boldsymbol{R}) \int \frac{\partial m_z}{\partial \rho}\rho^2 d\rho,
\end{equation}
where $\nabla_{\perp}=\frac{\partial}{\partial x}\hat{\boldsymbol{x}}+\frac{\partial}{\partial y}\hat{\boldsymbol{y}}=\frac{\partial}{\partial r}\hat{\boldsymbol{r}}+\frac{1}{r}\frac{\partial}{\partial \varphi}\hat{\boldsymbol{\varphi}}$ is the in-plane gradient. 

\section{Evaluation of the Force Integral \texorpdfstring{$\mathcal{I}$}{I}} \label{app:eq10}

\begin{figure}[!t]
\centering
\includegraphics[width=0.8\linewidth]{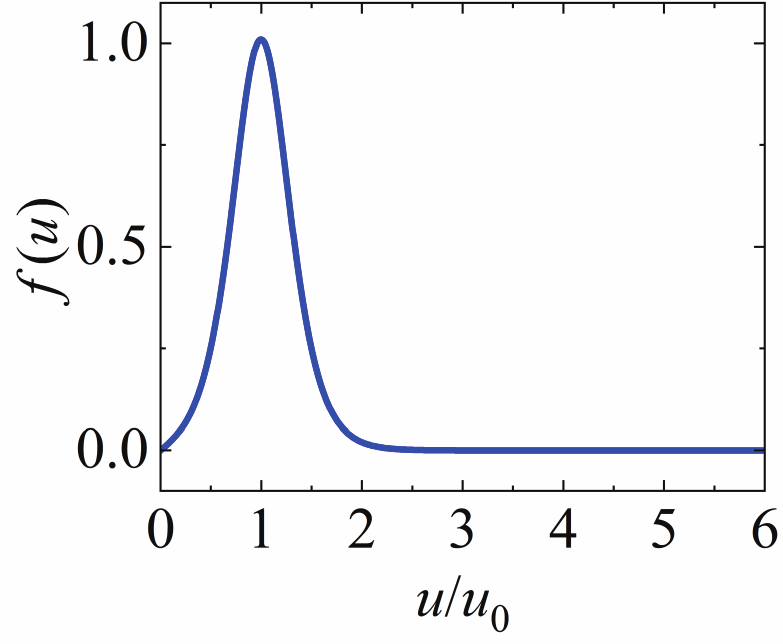}
\caption{The function $f(u)$ of Eq. \ref{fu} as a function of $u/u_0$.}
\label{fu-a1}
\end{figure}

The integral $\mathcal{I}$ defined in Eq.~\eqref{integral} is central to calculating the magnitude of the IFE-induced force. Using the parametrization $u = \rho/r_d$ and $u_0 = R_d/r_d$, it reads
\begin{equation}
\mathcal{I} = r_d^2 \int_0^{\infty} \frac{4 \sinh(u) \cosh(u) \sinh^2(u_0)}{[\sinh^2(u_0) + \sinh^2(u)]^2} \, u^2 \, du.
\label{integral_full}
\end{equation}
For a typical magnetic skyrmion, the ratio $u_0 = R_d / r_d$ characterizes its spatial structure, which is sufficiently greater than 1 to ensure that the function 
\begin{equation}
f(u) = \frac{4 \sinh(u) \cosh(u) \sinh^2(u_0)}{[\sinh^2(u_0) + \sinh^2(u)]^2}, 
\label{fu}
\end{equation}
is sharply peaked around $u = u_0$ as shown in Fig. \ref{fu-a1}. This allows for a peak-width approximation to evaluate the integral analytically and obtain a simple scaling relation.

We rewrite the integrand by noting that for $u_0 > 1$, $\sinh^2(u_0) \approx \frac{1}{4}e^{2u_0}$ dominates over $\sinh^2(u)$ except near $u \approx u_0$. Introducing the shifted variable $x = u_0 - u$, the dominant term in the denominator becomes $\sinh^2(u_0) + \sinh^2(u) \approx \frac{1}{4}e^{2u_0}(1 + e^{-2x})$. Substituting and simplifying, we can obtain
\begin{equation}
\mathcal{I} \approx 4 r_d^2 \int_{-\infty}^{u_0} \frac{e^{2x}}{(1 + e^{2x})^2} \, (u_0 - x)^2 \, dx.
\label{integral_approx}
\end{equation}
The kernel $K(x) = e^{2x}/(1+e^{2x})^2$ is symmetric, normalized $\int_{-\infty}^{\infty} K(x) dx = \frac{1}{2}$, and decays exponentially for $|x| \gg 1$. Therefore, $(u_0 - x)^2 \approx u_0^2$ can be factored out of the integral over the narrow region where $K(x)$ is non-negligible. The decision to extend the lower limit of integration to negative infinity results in negligible error due to the exponential decay of $K(x)$. This approach allows for further analysis and precision in our calculations, leading to the following conclusions
\begin{equation}
\begin{aligned}
\mathcal{I} &\approx 4 r_d^2 u_0^2 \int_{-\infty}^{\infty} \frac{e^{2x}}{(1+e^{2x})^2} \, dx 
\\& = 2 (r_d u_0)^2 = 2 R_d^2.
\label{integral_result}
\end{aligned}
\end{equation}

This result, $\mathcal{I} \approx 2R_d^2$, reveals that the effective force primarily couples to the skyrmion's characteristic area $\sim \pi R_d^2$. The approximation leverages the fact that the magnetization gradient $\partial m_z/\partial \rho$ is significant only within the skyrmion's domain wall region located near $\rho \sim R_d$. The key condition for the validity of the approximation is that the skyrmion's core radius is distinct from its wall width, which is satisfied for typical isolated Néel skyrmions in chiral films. While a fully numerical evaluation of Eq.~\eqref{integral_full} yields a coefficient close to 2, the analytical approximation correctly captures the essential scaling $\mathcal{I} \propto R_d^2$, which is crucial for understanding the parametric dependencies of the GHz microwave force discussed in the main text.

\nocite{*}
\bibliography{aip-refs}

\end{document}